\documentstyle[prb,preprint,aps]{revtex}
\begin{document}
\title {Volatility and Agent Adaptability\\ in a
Self-Organizing Market } 
\author{N.F. Johnson$^1$, S. Jarvis$^1$,  R. Jonson$^1$, P.
Cheung$^2$, Y.R. Kwong$^2$, and P.M. Hui$^2$} %
\address {$^1$ Physics Department, Oxford  University, Oxford,
OX1 3PU, England} 
\address {$^2$ Physics Department, The Chinese University of
Hong Kong, Shatin, Hong Kong}  
%
\date{\today}
\maketitle

\begin{abstract} We present results for the so-called
`bar-attendance' model of market behavior: $p$ adaptive agents,
each possessing $n$ prediction rules chosen randomly from a
pool, attempt to attend a bar whose cut-off is $s$. The global
attendance time-series has a mean near, but not equal to, $s$.
The variance, or `volatility', can show a {\em minimum} with
increasing adaptability of the individual agents.


\end{abstract}  
\newpage The dynamical properties of complex adaptive systems
are beginning to attract significant attention in many
disciplines
\cite{Casti,holland}. Economics, which historically was based
on notions of static equilibria involving rational agents, is
now the subject of much of this attention. Of particular
interest in  the field of Finance are the {\em microscopic}
factors which can give rise to {\em macroscopic} market
fluctuations, or `volatility'
\cite{everyone,stochastic,Challet,Savit}. 

Arthur \cite{Arthur} has proposed the so-called `bar-attendance'
model to investigate the global behavior in a market containing
heterogeneous agents with bounded rationality acting via
inductive reasoning. Specifically $p$ adaptive agents, each
possessing $n$ prediction rules or `predictors' chosen randomly
from a pool of $N$, attempt to attend a bar, whose cut-off is
$s$, on a particular night each week. Each week the agents
update their best rule for predicting a given week's attendance
based on the past attendance time-series $x(t)$. This feedback
mechanism with its adaptive feature provides an essential
ingredient for creating complex dynamics of $x(t)$. A
`mean-field' solution for agent behavior, whereby the majority
of agents use a given predictor in a given week, is unstable
since it will lead to a large deviation of $x(t)$ from the
cut-off $s$ for that week. Other features of interest to a
physicist include the fact that the problem lies in the
mesoscopic regime in terms of particle (agent) number  (i.e.
$p\sim 10^2$), the fact that the interactions between particles
(agents) are non-local in time and space, and the fact that the
basic bar model's evolution is purely deterministic
\cite{stochastic}.

Here we present results of extensive computer simulations for
the bar-attendance model.  We find that the volatility of the
attendance time-series $x(t)$ can show a {\em minimum} at
small, but finite, $n$. Hence increasing agent adaptability
does not generally lead to lower market volatility: in
particular it typically {\em increases} market volatility,
thereby contradicting the idea that
well-developed markets with `expert' traders should
be inherently less volatile than emerging markets. 

The computer model setup is as follows. The pool of $N$
predictors is chosen to  encompass a variety of simple, yet
realistic, prediction rules. Unlike the recent works of Challet 
and Zhang 
\cite{Challet} and Savit et al. \cite{Savit}, we do not
restrict all predictors to depend on the same number of past
weeks' data.  The predictors in our pool of $N$ are chosen from
a variety of `classes' of rule: one class might comprise rules
which take an arithmetic (class (i)), geometric (class
(ii)),  or weighted (class (iii)) average over the past $m$
weeks' attendances; rules from another class (class (iv)) 
might copy the result from week $m'$; alternatively, the mirror
image of $x(t)$ about $s$ might be taken from week $m''$ (class
(v)).  Allowing $m$, $m'$ and $m''$ to vary from unity to a few
tens, for example, would generate a pool containing $N\approx
400$ rules. We have found that the results of our simulations
are quite general provided that a variety of rule-classes are
always represented in the pool; this is reasonable since, in
real markets, professional analysts will often have quite
different recipes for predicting future trends. At the start of
a given simulation, each agent chooses $n$ rules from the pool
of $N$ (repetitions are allowed) and an arbitrary sequence of
attendances is input to simulate earlier weeks' data. We have
checked that the long-time behaviour, as indicated  by the
convergence of the standard deviation of attendance,  of
the model is generally insensitive to the initial
attendance-string. Apart from the predictors, there is an
additional difference between the present model and that of
Refs 5 and 6: the global parameter representing the weekly
outcome in the present model is the actual attendance number,
as opposed to a simple binary digit. This allows us to analyse
microscopic conditions before an attendance `crash' or
`bubble'. Such signatures are discussed elsewhere \cite{Us}. 

We have studied several weekly update schemes for our basic bar
model.  One scheme (method I) involves each agent
choosing from his $n$ predictors based on the cumulative performance 
of each of these $n$ predictors: when a given 
week's attendance $x_{m}$ is known, each agent examines each 
of his $n$ predictors to see which, in hindsight, would have 
worked.  A cumulative performance can then be assigned to
each  predictor according {\em either} to  the decision (i.e. go or
stay) implied by the outcome {\em or} to the error by which the
predictor's outcome differed from $x_{m}$.  The  predictor with the best
cumulative performance is then used to  decide the action to be taken
in week $m+1$.  This  scheme introduces a certain `reluctance to
change' -- i.e. a  predictor with a good track record will not be
rejected just  because of one bad performance. 
An alternative scheme (method II) involves the
best rule  for last week's data being used to predict the outcome of
the coming week: when a given week's attendance $x_m$
is known, each agent examines each of his $n$ predictors to see
which, in hindsight, would have worked best. He then uses this model to
predict the outcome for week $m+1$. If the predicted outcome is greater
than $s$, he stays at home. If it is less than $s$, he attends the
bar. If he attends when the actual number attending is  less
than $s$, he is satisfied. If it is more, he is dissatisfied. 
We have carried out simulations using both methods
and the  results are qualitatively similar with some minor
differences; in particular, method I gives a distribution  of weekly
attendance with a sharper peak than method II. 

The individual agents' perspective provides a local utility -- a
given agent is happy with any attendance smaller than $s$ as
long as he is one of the attendees. However, large fluctuations
represent a wasteage: a negative (positive) fluctuation away
from $s$ in any given week implies that he should have attended
(stayed away). The individual agents are analogous to traders
deciding a day's trading strategy ahead of time, e.g. setting up
the computer trades for a given day. For example, imagine each
trader (i.e. agent) wishes to buy a given quantity of a
particular currency (i.e. he wishes to attend the bar) on a
daily  basis. Using a simple argument of supply and demand, the
price of the currency will rise according to the number of
traders (agents) attempting to buy that currency (attend the
bar). On a given day, therefore, if too many other traders
(agents) try to buy currency (attend the bar) then it will push
the price (attendance) above the desirable cutoff value $s$. A
large volatility will imply large risk for the trader.  The
bar-manager has a global utility: for a given mean attendance
$\bar x$, he will  wish to keep the attendance fluctuations
(volatility) as small as possible so that he can plan staffing,
beer orders etc. He does not, however, care whether an
individual agent attends or not. The bar-manager is hence
analogous to a government who wishes  for a low exchange-rate
volatility. A low volatility
might therefore be preferable both locally and globally. A
detailed analysis of utilities and gains/losses by individual
agents are discussed elsewhere \cite{Us}. Here we concentrate
on the behavior of  the  volatility $\Delta
x=\sqrt{\langle(x(t)-{\bar x})^2\rangle}$ as a function of $n$,
$p$, $s$ and $N$. 

Figure 1 shows a plot of the volatility $\Delta x$ as a function
of the number of predictors $n$ per agent, for the basic bar
model.  The pool of predictors consists of $N=45$ rules  from
classes (i)-(iv), and method I is used for the  weekly
updating.  There is a minimum in
$\Delta x$ for $n\sim 4$. Note that this minimum remains even
after `configurational averaging' (i.e. averaging over an
ensemble of simulations with different initial
attendance-strings and different distributions  of rules among
agents).  It is also robust against the precise rules contained
in the pool $N$, provided that various rule-classes are present
as discussed above.  The minimum persists  when rules of class
(v) are also included.   The reason for a minimum can be
understood as follows. For $n=1$, each agent has no
adaptability; even if he is initially dealt a poorly-performing
predictor, he has no option but to repeatedly apply it.
Increasing $n$ gives him the possibility of replacing  this
predictor with a better one; $\Delta x$ then decreases with
increasing $n$. However now consider the limit $n\rightarrow N$;
the agents will hold, on average, increasingly similar toolbags
of predictors.  In particular, the total attendance will tend
to avoid the value $s$, instead forming peaks on either side.
The volatility will be large. A minimum in $\Delta x$ at small
$n$ is therefore reasonable.  The choice of the
values of the parameters implies  an appreciable overlap of the
$n$ rules among the toolbags of the agents: however the overlap
is not large enough to cause identical decisions
by the agents.  It should be noted  that for given $n$,
$s$, and $p$, $\Delta x$ decreases as $N$  increases and
eventually saturates to a finite large-$N$ limit.  

Figure 2 shows a plot of the volatility $\Delta x$ and mean
$\bar x$ as a function of the number of agents $p$ for two 
different values of $N$.    For $p<s$, $\bar x\approx p$ and
$\Delta x\approx 0$ as expected. For $p>s$, the mean
attendance  is similar, but not equal, to $s$. On closer
inspection we note that for $s<p<2s$, $\bar x$ is slightly less
than $s$ while for $p>2s$, $\bar x$ becomes greater than $s$.
The volatility $\Delta x$ seems to follow a square-root-like
dependence on $p$ for $p>s$. 

It is tempting to suggest a  random walk argument in order to
explain the results in Fig. 2. Suppose that the attendence is a
stochastic process: specifically imagine it is a first-order
Markov process so that the past weeks' data does not contain
any extra information for the agent. To compensate for this
loss of information, we will supply him with an extra piece of
information by giving him the value of $p$ (N.B. In the basic
bar model simulation,  we did {\em not} supply each agent with
the value of $p$). Given the common knowledge of the cutoff
$s$, each agent will attend with a probability $\frac{s}{p}$
and stay away with a probability $1-\frac{s}{p}$. Each of the
$p$ agents carries out the same calculation, hence the average
weekly attendance (i.e.
$\bar x$) should be $p \cdot  \frac{s}{p}=s$. Following standard
random walk results\cite{Reif}, the corresponding volatility 
$\sim\sqrt{4  \cdot  p  \cdot  \frac{s}{p}  \cdot 
(1-\frac{s}{p})}$, i.e.
$\Delta x\sim \sqrt{4s(1-\frac{s}{p})}$. The solid line
on Fig. 2 shows that this is a fairly good approximation. In
fact it is better than expected. It is only in the large-$N$
limit that the stochastic, random walk limit should actually be
quantitatively reasonable: in the large-$N$ limit the pool is so
large that agents are highly unlikely to have any rules in
common. The agents therefore are effectively no longer competing
against each other in terms of predictors. On closer inspection
of this random walk argument, one finds that while  the
$\sqrt{s(1-\frac{s}{p})}$ factor persists, the factor 4 in the
expression for $\Delta x$ is not justified given that the walk
is confined to the half-space $x\geq 0$; hence the random walk
scaling should only be regarded as a qualitative guide. An
alternative random walk model can be obtained as follows. Each
agent regards the predicted outcome from his set of $n$
predictors as being so complex that he might as well toss a
coin when deciding whether to attend in a given week. The
probability for attendance is hence $0.5$, leading to $\bar
x\sim
\frac{p}{2}$ and $\Delta x\sim \sqrt{p}$. Although this model
captures the steady rise in $\bar x$ with $p$ observed in Fig.
2, together with the $\sqrt{p}$-like dependence of $\Delta x$,
it is also only useful as a qualitative guide. We have not been
able to find a stochastic model which reproduces the
quantitative results of the simulations -- this is perhaps not
surprising given the complexity of the dynamical system being
studied.

Figure 3 shows a plot of the volatility $\Delta x$ as a function
of cutoff $s$ for $0<s<p$. The qualitative 
scaling  $\sqrt{4s(1-\frac{s}{p})}$ based  on random
walk arguments is also shown.   
\vskip0.2in

Finally, we  mention two generalizations of the basic
bar model which can be included in an attempt to capture
additional features of real agent behavior. A full discussion 
and results are presented elsewhere \cite{Us}.  \vskip0.1in

\noindent (i) Evolution and Learning.

\noindent Instead of retaining the same $n$ predictors
throughout the simulation, each agent will throw back into the
pool a predictor which persistently underperforms. He will then
randomly choose another predictor from the pool. In this way,
predictors can become either `live' or `dead' during the
simulation. Since the pool is of fixed size $N$, the situation
can arise where a number of agents begin to use the same
successful predictor, hence nullifying its accuracy. This may
lead to a `mass-killing' of this predictor; simultaneously, it
allows a predictor that had lain `dead' in the pool the
possibility of being resurrected during the random picking
process. Hence we have introduced a degree of evolution and
learning into each agent's predictor-set.

\vskip0.1in \noindent (ii) Irrational behaviour in a crowd.

\noindent In the basic model, each agent tries to profit from
the accuracy of his best model (i.e. the model with minimum
error). Here, however, we introduce the feature whereby a small
subset of the $N$ predictors are evaluated inaccurately. All
agents holding the affected predictors will therefore be misled
by the outcomes. This feature is reminiscent of how a rumour
spreading through a given subset of traders can affect their
decisions and, ultimately, the markets themselves. We find that
this feature can give rise to  intermittency effects in the
volatility \cite{Us}. 

\vskip0.3in In summary we have presented an analysis of the
bar-attendance model. This model offers a simple paradigm for a
competitive marketplace where agents with bounded rationality
act using inductive reasoning. We hope that the present work
will stimulate further interest in what is proving to be an
exciting field of study for physicists.   

\vskip0.3in One of us (N.F.J.) would like to thank Colin Mayer,
David Sherrington, Philippe Binder, Robert Bacon and
Jean-Philippe Bouchaud for useful discussions at various stages
during this work.  This work was supported in part by a grant 
from the British Council and the Research Grants Council of 
the Hong Kong SAR Government through the UK-HK Joint Research
Scheme 1998.

\newpage \centerline{\bf Figure Captions}

\bigskip

\noindent Figure 1: Volatility $\Delta x$ as a function of the
number of predictors $n$ per agent. $N=45$, $p=100$ and
$s=60$.  
\bigskip 

\noindent Figure 2: Volatility $\Delta x$ and mean $\bar x$ as a
function of the number of agents $p$. $n=5$, $s=60$ and $N=45$
and $400$. The random walk
scaling $\sqrt{4s(1-\frac{s}{p})}$ is also shown. \bigskip

\noindent Figure 3: Volatility $\Delta x$ as a function of the
cutoff $s$. $n=5$, $s=60$ with $N=45$ and $400$. The random walk
scaling $\sqrt{4s(1-\frac{s}{p})}$ is also shown (dashed line).

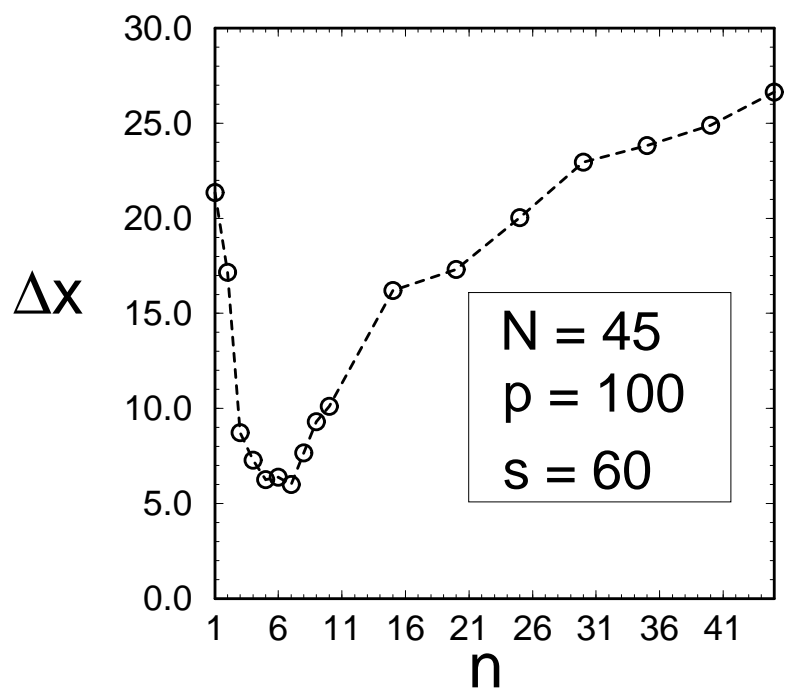
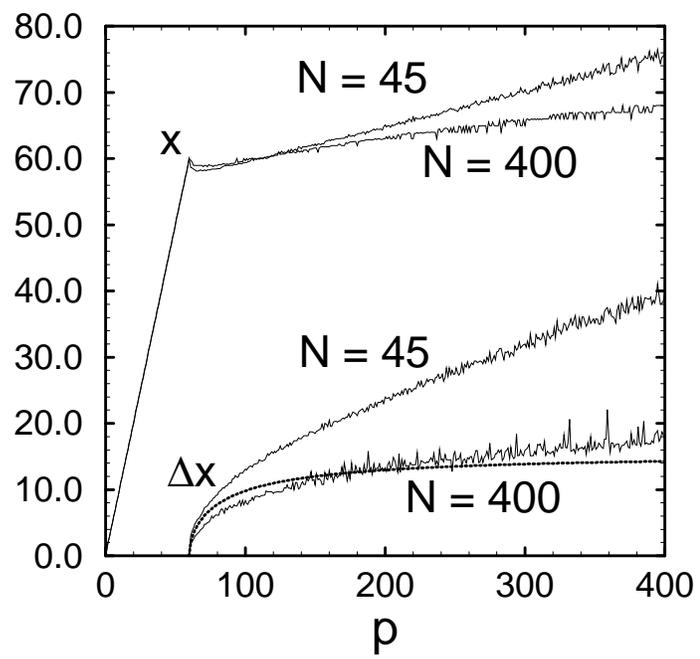
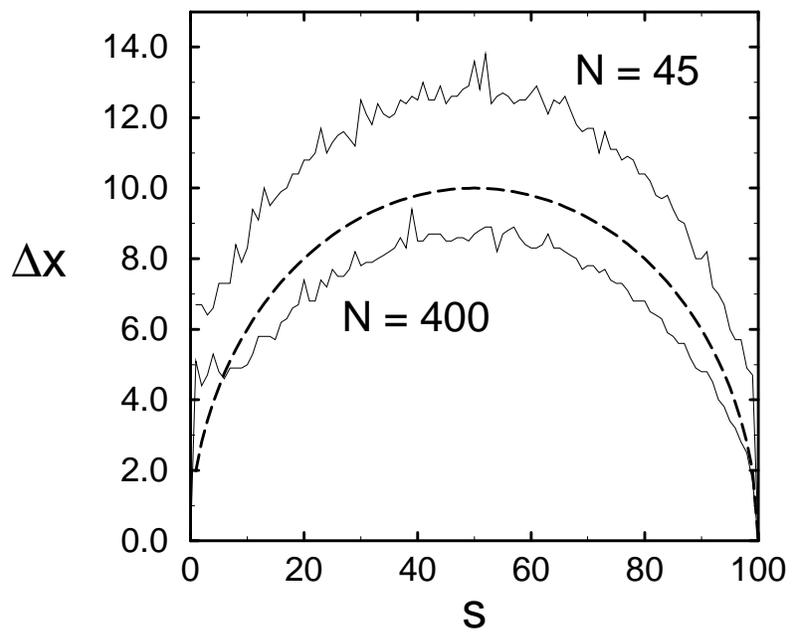
\end{document}